\documentclass[aps,prd,twoside,twocolumn,nofootinbib,10pt,superscriptaddress,%
showpacs,floatfix]{revtex4-1}

\usepackage{amsmath,amssymb}
\usepackage{graphicx,bm}
\usepackage{slashed}
\usepackage{dcolumn}
\usepackage{epstopdf}
\usepackage{ulem} 
\usepackage[usenames]{color}
\usepackage{subfigure}
\usepackage{float}
\usepackage[usenames]{color}

\allowdisplaybreaks

\renewcommand\sout{\bgroup \color{red} \ULdepth=-.5ex \ULset}

\begin{document}

\title{\boldmath ${}_{\Lambda\Lambda}^{\ \, 4}$H in halo effective field theory}

\author{Shung-Ichi Ando}
\email{sando@daegu.ac.kr}
\affiliation{Department of Physics Education, Daegu University, Gyeongsan, 
Gyeongbuk 712-714, Korea}

\author{Ghil-Seok Yang}
\affiliation{Department of Physics, Kyungpook National University, 
Daegu 702-701, Korea}

\author{Yongseok Oh}
\email{yohphy@knu.ac.kr}
\affiliation{Department of Physics, Kyungpook National University, 
Daegu 702-701, Korea}
\affiliation{Asia Pacific Center for Theoretical Physics, Pohang, 
Gyeongbuk 790-784, Korea}

\date{}
\begin{abstract}
The ${}_{\Lambda\Lambda}^{\ \ 4}$H bound state and the $S$-wave 
hypertriton(${}_\Lambda^{ \, 3}$H)-$\Lambda$ scattering in spin singlet and 
triplet channels below the hypertriton breakup momentum scale are studied in 
halo/cluster effective field theory at leading order by treating the 
${}_{\Lambda\Lambda}^{\ \ 4}$H system as a three-cluster 
($\Lambda$-$\Lambda$-deuteron) system.
In the spin singlet channel, the amplitude is insensitive to the cutoff parameter 
$\Lambda_c$ introduced in the integral equation, and we find that there is no 
bound state. In this case, the scattering length of the hypertriton-$\Lambda$ 
scattering is found to be $a_0^{} = 16.0\pm 3.0$~fm.
In the spin triplet channel, however, the amplitude obtained by the coupled integral 
equations is sensitive to $\Lambda_c$, and we introduce the three-body contact 
interaction $g_1^{} (\Lambda_c)$.
After phenomenologically fixing $g_1^{} (\Lambda_c)$, we find that 
the correlation between the two-$\Lambda$ separation energy $B_{\Lambda\Lambda}$
from the $_{\Lambda\Lambda}^{\ \ 4}$H bound state and 
the scattering length $a_{\Lambda\Lambda}^{}$ of the $S$-wave $\Lambda$-$\Lambda$
scattering is significantly sensitive to the value of $\Lambda_c$. 

\end{abstract}


\maketitle

\section{Introduction}
\label{sec:Introduction}

Light double-$\Lambda$ hypernuclei are exotic few-body systems that provide
opportunities to investigate the flavor SU(3) structure of baryon-baryon 
interactions in the strangeness $\mathcal{S}=-2$ channel~\cite{RSY99,PHM07,FSN06}.
They are also expected to have a key role to resolve the long-standing puzzle on
the existence of the $H$-dibaryon~\cite{Jaffe77}, which attracts recent interests
triggered by lattice QCD simulations~\cite{NPLQCD-10,HALQCD-10}.
Since the seminal experiments on double-$\Lambda$ hypernuclei of
Refs.~\cite{DGPP63,DDFM89}, however, there are only a few reports on the 
observation of double-$\Lambda$ hypernuclei and, as a result, our understanding 
on these systems is still very poor.
In the KEK-E373 experiment the $\Lambda\Lambda$ interaction energy%
\footnote{The $\Lambda\Lambda$ interaction energy $\Delta B_{\Lambda\Lambda}$ is
defined as $\Delta B_{\Lambda\Lambda} ({}_{\Lambda\Lambda}^{\ \, A}Z) = 
B_{\Lambda\Lambda} ({}_{\Lambda\Lambda}^{\ \, A}Z) - 2 B_{\Lambda} ({}_{\ \ \
\Lambda}^{A-1}Z)$, where $B_{\Lambda\Lambda}$ and $B_\Lambda$ are binding 
energies of the corresponding nuclei.}
inside ${}_{\Lambda\Lambda}^{\ \ 6}$He is measured as 
$\Delta B_{\Lambda\Lambda} \simeq 1.0$~MeV, which suggests a weakly attractive 
$\Lambda\Lambda$ interaction~\cite{TAAA01}.
In addition, the formation of another double-$\Lambda$ hypernucleus,
${}_{\Lambda\Lambda}^{\ \ 4}$H, is conjectured in the BNL-AGS E906 
experiment~\cite{AAAB01}.
Theoretically, although the first Faddeev-Yakubovsky calculation showed a
negative result~\cite{FG02}, subsequent theoretical 
studies~\cite{NAM02,Shoeb04,NSAM04,SUB11}
predicted the possibility of the ${}_{\Lambda\Lambda}^{\ \ 4}$H bound state
based on the phenomenological $\Lambda\Lambda$ potentials which can describe 
the bound state of ${}_{\Lambda\Lambda}^{\ \ 6}$He.

Since the stability of double-$\Lambda$ hypernuclei depends on the 
$\Lambda\Lambda$ interaction, more accurate information on this interaction is 
strongly required.
Recently, the scattering length $a_{\Lambda\Lambda}^{}$ of $S$-wave 
$\Lambda\Lambda$ scattering is deduced from the 
${}^{12}$C($K^-,K^+\Lambda\Lambda X)$ 
reaction~\cite{YAAF07}, which leads to 
$a_{\Lambda\Lambda}^{} = -1.2 \pm 0.6$~fm~\cite{GHH11},
and the data for the Au+Au collisions at the Relativistic Heavy Ion Collider~\cite{STAR13} 
are analyzed to obtain $a_{\Lambda\Lambda}^{} \ge -1.25$~fm in Ref.~\cite{ExHIC-13}.
These values are consistent with those extracted from the leading order 
calculations for the $\mathcal{S} = -2$ baryon-baryon interactions in 
chiral effective theory~\cite{PHM07} and in the Nijmengen ESC04d 
phenomenological potential model~\cite{RY06b}.
On the other hand, other phenomenological potential model predictions are scattered in values 
from $-0.27$~fm to $-3.804$~fm even though such models could explain the existence of the 
${}_{\Lambda\Lambda}^{\ \ 6}$He bound state.
The present situation is summarized, for example, in Table~I of Ref.~\cite{GHH11}.
This may imply that the parameter space of potential models would be too large to determine 
unambiguously the parameter values from the currently available experimental data. 
In such a situation, it would be worth studying the structure of hypernuclei by employing 
a very low energy effective field theory (EFT) which has a low separation scale, a well-defined 
expansion scheme, and a few parameters to determine.

The methods of EFT nowadays become popular in many fields. 
(For a review, see, e.g., Refs.~\cite{BV02c,BH04}.)
In this scheme, a theory is constructed based on a scale which separates 
low energy and high energy degrees of freedom, and the theory constructed 
in such a way provides a systematic perturbative expansion in powers of 
$Q/\Lambda_H$, where $Q$ is the typical scale of the reaction in question and 
$\Lambda_H$ is the large (or high energy) scale.
High energy degrees of freedom above $\Lambda_H$ are integrated out and their 
effects are accounted for through the coefficients of contact interactions, 
so-called low energy constants, in higher order.

In this work, we investigate the relation between the ${}_{\Lambda\Lambda}^{\ \ 4}$H 
bound state and the $S$-wave hypertriton-$\Lambda$ scattering below the 
hypertriton breakup momentum for spin singlet and triplet channels by employing 
Halo/Cluster EFT at leading order (LO).
In particular, we treat the ${}_{\Lambda\Lambda}^{\ \ 4}$H hypernucleus
as a three-body $\Lambda\Lambda d$ system, where $d$ stands for a deuteron.
Although the scattering experiment with double-$\Lambda$ systems 
is not feasible in near future, a qualitative/theoretical information from the scattering results 
can be possibly connected to the bound state problem, which makes the main motivation
of the present work.

Below the hypertriton breakup momentum, we can choose the typical momentum ($Q$) 
of the reaction as the $\Lambda$ particle separation momentum from the hypertriton,
which is defined by $\gamma_{\Lambda d}^{} = \sqrt{2\mu_{\Lambda d}^{} B_\Lambda} 
\simeq 13.5\pm 2.6$~MeV, 
where $\mu_{\Lambda d}^{}$ is the reduced mass of the $\Lambda d$ system 
and $B_\Lambda$ is the $\Lambda$ particle separation energy 
from the hypertriton, $B_\Lambda^{\rm expt.} \simeq 0.13\pm 0.05$~MeV~\cite{JBKK73}.
On the other hand, the large (high momentum) scale $\Lambda_H$ 
is chosen to be the deuteron binding momentum, 
$\gamma =\sqrt{m_N^{} B_2}\simeq 45.7$~MeV, where $m_N^{}$ is the 
nucleon mass and $B_2$ is the deuteron binding energy, $B_2\simeq 2.22$~MeV.   
Then our expansion parameter is 
$Q/\Lambda_H\sim \gamma_{\Lambda d}^{}/\gamma \simeq 1/3$, which supports
our expansion scheme. 
Because the deuteron is not broken up into two nucleons at low momentum
below the deuteron binding momentum, we may treat the deuteron field 
as a cluster field, i.e., like an elementary field.

The $\Lambda\Lambda d$ system can form spin singlet and spin triplet states for
the ${}_{\Lambda\Lambda}^{\ \,4}$H channel and we consider only the $S$-wave
case for the relative orbital angular momentum.
For the spin singlet channel of the $S$-wave hypertriton-$\Lambda$ scattering, 
we obtain a single integral equation for the scattering amplitude, which is parameterized 
by the effective range parameters of the $S$-wave $\Lambda$-$d$ scattering 
in the hypertriton channel, namely, the scattering length $a_{\Lambda d}^{}$ 
(or equivalently the hypertriton binding momentum $\gamma_{\Lambda d}^{}$) 
and the effective range $r_{\Lambda d}^{}$. 
The integral is regularized by introducing a sharp momentum cutoff $\Lambda_c$ 
in the integral equation.
We find that when the cutoff $\Lambda_c$ is larger than $\Lambda_H$, there is 
no cutoff dependence in the results, which implies that the system is insensitive to 
the short range mechanism~\cite{Griess05}.
This then suggests that introducing a three-body contact interaction at LO is not
necessary.
In addition, here we employ the standard Kaplan-Savage-Wise (KSW) counting rules~\cite{KSW98a},
where the effective range, $r_{\Lambda d}^{}$, is treated as a higher order term.
This shows that the scattering length $a_0^{}$ and the phase shift $\delta_0^{}$ of the $S$-wave 
hypertriton-$\Lambda$ scattering are well controlled by $\gamma_{\Lambda d}^{}$.

On the other hand, for the spin triplet channel, coupled integral equations are obtained 
for the scattering amplitudes.  
Because of spin statistics these equations consist of two cluster channels. 
One is the hypertriton-$\Lambda$ channel of spin-1 and the other is the deuteron and 
double-$\Lambda$ system, 
where we assume that the double-$\Lambda$ is described by the $\Lambda\Lambda$-dibaryon
state and the components in the cluster states are in relative $S$-wave.
We find that the coupled integral equations show a sensitivity to the cutoff $\Lambda_c$.
Thus, as in the case of three-nucleon system in the triton channel within pionless 
EFT~\cite{BHV99}, a three-body contact interaction needs to be introduced in order to make 
the results cutoff-independent.
In addition, within the standard KSW counting rules~\cite{KSW98a} 
the dressed composite propagators of the hypertriton for the $\Lambda$-$d$ composite 
state and of the dibaryon for two $\Lambda$ particles in ${}^1S_0$ state are expanded 
in terms of the effective range parameters. 
Thus the coupled integral equations are represented in terms of only four parameters 
at LO, namely, $\gamma_{\Lambda d}^{}$, $a_{\Lambda\Lambda}^{}$, the coupling of the 
three-body contact interaction $g_1^{} (\Lambda_c)$, and the cutoff $\Lambda_c$. 
Unlike the effective range parameters, however, there are no experimental data to  
constrain $g_1^{} (\Lambda_c)$ for ${}_{\Lambda\Lambda}^{\ \,4}$H.

Because of the paucity of empirical information to constrain the low energy 
constants it is very hard to draw a robust prediction
on the existence of the bound state in the ${}_{\Lambda\Lambda}^{\ \,4}$H channel.
Therefore, instead of tackling the problem on the existence of bound states 
we investigate the effect of the contact term in the ${}_{\Lambda\Lambda}^{\ \,4}$H
system.
For this purpose we consider two cases.
In the first case, we do not include the contact interaction by setting
$g_1^{} = 0$.
Then the system is found to have a large negative scattering length at 
$\Lambda_c \simeq \Lambda_H$, which may imply the formation of a quasi-bound state.
Furthermore, if $\Lambda_c$ is sent to the asymptotic limit, $\Lambda_c \to \infty$, 
we find that a bound state arises in the system.

In the second case, we turn on the contact interaction.
To constrain the value of $g_1^{}(\Lambda_c)$, 
we employ the results of the potential model calculations of 
Refs.~\cite{FG02,NAM02} and determine $g_1^{}(\Lambda_c)$ by using
the computed double-$\Lambda$ separation energy 
$B_{\Lambda\Lambda}$ of ${}_{\Lambda\Lambda}^{\  \ 4}$H for given values of 
$a_{\Lambda\Lambda}^{}$.
Then we find that the renormalized $g_1^{} (\Lambda_c)$ exhibits so-called 
the limit-cycle when $\Lambda_c$ is sent to the asymptotic limit. 
In the present work, we also calculate $B_{\Lambda\Lambda}$ as a function of 
$a_{\Lambda\Lambda}^{}$ for a fixed $g_1^{}(\Lambda_c)$ and a correlation between 
$B_{\Lambda\Lambda}$ and $1/a_1^{}$ as well, where $a_1^{}$ is the scattering length
of the $S$-wave hypertriton-$\Lambda$ scattering in the spin triplet channel at LO.
We find that the $a_{\Lambda\Lambda}^{}$-dependence of $B_{\Lambda\Lambda}$ 
is quite sensitive to the value of $\Lambda_c$.
For example, $B_{\Lambda\Lambda}$ is found to be almost insensitive 
to $a_{\Lambda\Lambda}^{}$ when $\Lambda_c\simeq \Lambda_H$.
On the other hand, the reported $a_{\Lambda\Lambda}^{}$-dependence of 
$B_{\Lambda\Lambda}$ in the potential model calculations of 
Refs.~\cite{FG02,NAM02} is recovered when $\Lambda_c\simeq 6 \Lambda_H$.
In the present work, we will investigate the implications of the choice on the 
cutoff $\Lambda_c$ and the $a_{\Lambda\Lambda}^{}$- and $\Lambda_c$-dependence 
of the properties of ${}_{\Lambda\Lambda}^{\ \,4}$H system in the cluster theory.

This paper is organized as follows.
We start with the relevant effective Lagrangian in the next section, which
defines notations and our basic tools for studying hypernuclei.
In Sec.~\ref{sec:2body}, the two-body parts of the $\Lambda\Lambda d$ system,
i.e., the dressed $\Lambda\Lambda$ dibaryon propagator in ${}^1S_0$ channel and 
the dressed hypertriton propagator (as a $\Lambda d$ system), are constructed.
In Sec.~\ref{sec:3body}, the integral equations of the $\Lambda\Lambda d$ 
three-body system for the $S$-wave hypertriton-$\Lambda$ scattering are 
constructed in the spin singlet and triplet states.
The numerical results are presented in Sec.~\ref{sec:results} and 
Section~\ref{sec:discussion} contains a summary and conclusions of this work.


\section{Effective Lagrangian}
\label{sec:Lag}

In EFT, effective Lagrangian is constructed on the symmetry requirement with
relevant degrees of freedom at low energies being expanded in terms of the 
number of derivatives order by order~\cite{Wei79}. 
The effective Lagrangian at LO for this work can be written as
\begin{eqnarray}
\mathcal{L} =
\mathcal{L}_\Lambda + \mathcal{L}_d+ \mathcal{L}_{s}+ \mathcal{L}_{t}
+ \mathcal{L}_{\Lambda t} .
\label{eq:Lag}
\end{eqnarray}
Here, $\mathcal{L}_\Lambda$ and $\mathcal{L}_d$ are the standard one-body
$\Lambda$ and (elementary) deuteron Lagrangian in the heavy-baryon 
formalism~\cite{BKM95}, which read
\begin{eqnarray}
\mathcal{L}_\Lambda &=& \mathcal{B}_\Lambda^\dagger 
\left[ i v \cdot \partial +\frac{(v\cdot \partial)^2 - \partial^2}{2m_\Lambda^{}}
\right] \mathcal{B}_\Lambda 
+ \cdots ,
\\
\mathcal{L}_d &=& 
d_i^\dagger \left[ i v \cdot \partial +\frac{(v\cdot \partial)^2 - \partial^2}{2m_d^{}}
\right] d_i 
+ \cdots ,
\end{eqnarray}
where $\mathcal{B}_\Lambda$ is the $\Lambda$-baryon field of spin-1/2, $d_i^{}$ is the 
deuteron (vector) field of spin-1, and $v^\mu$ is a velocity vector with 
$v^\mu=(1,\bm{0})$ in our case.
The $\Lambda$ and deuteron masses are represented by $m_\Lambda^{}$ and
$m_d^{}$, respectively.
The dots denote the higher order terms that are irrelevant for the LO
calculations.

Equation~(\ref{eq:Lag}) also contains the Lagrangian for the composites
containing strangeness. 
For this purpose, we introduce $s$ and $t$ fields to denote the $\Lambda\Lambda$
dibaryon in the ${}^1S_0$ state and the $\Lambda d$ composite in the 
${}^2S_{1/2}$ state.
Then $\mathcal{L}_s$ and $\mathcal{L}_t$ are the Lagrangians for these fields
including $s \leftrightarrow \Lambda\Lambda$ and $t \leftrightarrow \Lambda d$
interactions, which read~\cite{BS00,AH04,AH12}
\begin{eqnarray}
\mathcal{L}_{s} &=& 
\sigma_s^{} s^\dagger \left[ i v\cdot \partial + \frac{(v\cdot \partial)^2-\partial^2}
{4m_\Lambda} + \Delta_s \right] s
\nonumber \\ && \mbox{} 
-y_s^{} \left[
s^\dagger \left( \mathcal{B}_\Lambda^T P^{(^1S_0)} \mathcal{B}_\Lambda \right) +\mbox{ H.c.}
\right] + \cdots ,
\\
\mathcal{L}_{t} &=& 
\sigma_t^{} t^\dagger \left[ iv\cdot\partial + \frac{(v\cdot\partial)^2 -\partial^2}
{2(m_d+m_\Lambda)}
+ \Delta_t \right] t
\nonumber \\ && \mbox{} 
+ \frac{y_t^{}}{\sqrt3}\left[ t^\dagger \vec{\sigma}\cdot \vec{d}\, \mathcal{B}_\Lambda +
\mbox{ H.c.}
\right] + \cdots ,
\end{eqnarray}
where $\sigma_{s}^{}$ and $\sigma_t$ are sign factors, 
$\Delta_{s}$ and $\Delta_t$ are the mass differences
between the composite states and their constituents, 
and $y_{s}^{}$ and $y_{t}^{}$ are coupling constants.
The spin projection operator of the $\Lambda\Lambda$ composite 
onto the $^1S_0$ state is 
\begin{eqnarray}
P^{(^1S_0)} = - \frac{i}{2}\sigma_2^{} ,
\end{eqnarray}

The three-body contact interaction is given by the Lagrangian $\mathcal{L}_{\Lambda t}$,
where $t$ and $\Lambda$ fields are in the $^3S_1$ channel, which reads
\begin{eqnarray}
\mathcal{L}_{\Lambda t} &=& - \frac{g_1^{}(\Lambda_c)}{\Lambda_c^2} 
\left( \mathcal{B}_\Lambda^T P^{(^3S_1)}_i t \right)^\dagger
\left( \mathcal{B}_\Lambda^T P^{(^3S_1)}_i t \right) 
+ \cdots,
\end{eqnarray}
with the spin projection operator onto the $^3S_1$ state,
\begin{eqnarray}
P^{(^3S_1)}_i = -\frac{i}{2}\sigma_2^{} \sigma_i^{}.
\end{eqnarray}
The coupling constant of the three-body contact interaction is given by 
$g_1^{}(\Lambda_c)$ as a function of the cutoff $\Lambda_c$ which will be 
introduced in the integral equations below.


\section{Two-body part}
\label{sec:2body}

\subsection{\boldmath $S$-wave $\Lambda\Lambda$ scattering in ${}^1S_0$ channel}

\begin{figure*}[t]
\begin{center}
\includegraphics[width=0.80\textwidth]{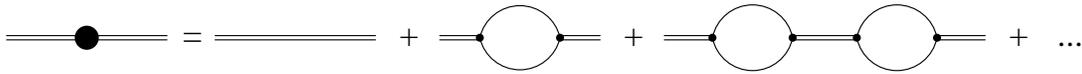}
\caption{
Diagrams for dressed dibaryon propagator.
In the right hand side, the double solid line represents the bare dibaryon propagator and 
the single solid line denotes the $\Lambda$ propagator.
}
\label{fig;dressed-propagator}
\end{center}
\end{figure*}

\begin{figure}[t]
\begin{center}
\includegraphics[width=0.4\columnwidth]{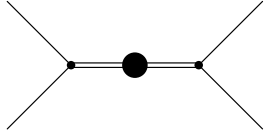}
\caption{
Diagram for $\Lambda\Lambda$ scattering amplitude.
A double line with a filled circle denotes a dressed 
propagator as explained in Fig.~\ref{fig;dressed-propagator}.
}
\label{fig;scattering-amplitude}
\end{center}
\end{figure}

At low energies, we assume that the dominant partial wave of $\Lambda\Lambda$ 
scattering is the ${}^1S_0$ state and the scattering process can be described 
by the effective range parameters.
Therefore, this is similar to the low-energy nucleon-nucleon scattering in the 
${}^1S_0$ channel studied, for example, in  Ref.~\cite{AH04}.
Diagrams for the dressed dibaryon field and for the scattering amplitude are 
shown in Figs.~\ref{fig;dressed-propagator} and \ref{fig;scattering-amplitude}, 
respectively.

Referring the details to Ref.~\cite{AH04}, we can obtain the scattering amplitude 
in the center-of-mass (CM) frame as
\begin{eqnarray} 
A(E) = \frac{4\pi}{m_\Lambda^{}}
\left( -\frac{1}{a_{\Lambda\Lambda}^{}} + 
\frac12 r_{\Lambda\Lambda}^{} k^2 - ik \right)^{-1},
\end{eqnarray}
where $a_{\Lambda\Lambda}^{}$ and $r_{\Lambda\Lambda}^{}$ are the scattering 
length and effective range of $\Lambda\Lambda$ scattering in the ${}^1S_0$ 
channel.
The on-shell total energy is $E=k^2/m_\Lambda^{}$ with $k = |\bm{k}|$.

Thus the renormalized dressed dibaryon propagator can be written as
\begin{widetext}
\begin{eqnarray}
D_{s}(p_0^{},\bm{p}) &=& \frac{4\pi}{m_\Lambda^{} y_s^2}
\left[
\frac{1}{a_{\Lambda\Lambda}} +\frac12r_{\Lambda\Lambda} 
\left( -m_\Lambda^{} p_0^{} + \frac14 \bm{p}^2 -i\epsilon \right)
- \sqrt{ -m_\Lambda^{} p_0^{} + \frac14 \bm{p}^2 -i\epsilon}
\right]^{-1}
\end{eqnarray}
\end{widetext}
and
\begin{eqnarray}
y_s^{} = -\frac{2}{m_\Lambda^{}}
\sqrt{ \frac{2\pi}{r_{\Lambda\Lambda}^{}} }  \ .
\end{eqnarray}
Here, $p_0^{}$ and $\bm{p}$ are the off-shell (loop) energy and momentum 
which do not satisfy the on-shell condition in the CM frame mentioned above. 
In addition, we have suppressed the cutoff dependence in the effective range parameters 
from the bubble diagrams. 
We use the same cutoff value for renormalizing $a_{\Lambda\Lambda}^{}$ 
and $r_{\Lambda\Lambda}^{}$ in the three-body part, 
which will be discussed in Sec.~\ref{sec:3body}.

\begin{figure*}[t]
\begin{center}
\includegraphics[width=0.80\textwidth]{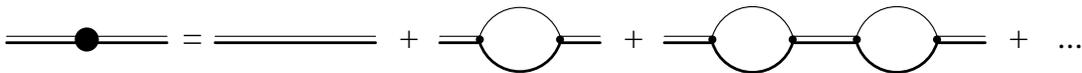}
\caption{
Diagrams for dressed hypertriton propagator as a $\Lambda d$ system.
In the right-hand side, the solid line denotes the $\Lambda$ hyperon while the
thick solid line represents the deuteron. 
The bare $t$ field as a $\Lambda d$ composite state in hypertriton channel is denoted by
the double (thin and thick) solid line. 
}
\label{fig;dressed-t-propagator}
\end{center}
\end{figure*}

\subsection{\boldmath $S$-wave $\Lambda d$ system in hypertriton channel}

The hypertriton (${}^{\,3}_\Lambda$H) has the quantum numbers of $J^\pi = 1/2^+$ 
and $T=0$, where $T$ stands for isospin,
and its $\Lambda$ separation energy is $B_{\Lambda} = 0.13 \pm 0.05$~MeV~\cite{JBKK73}. 
We refer the readers to Ref.~\cite{Hammer01} for a study on this state within pionless EFT.

Shown in Fig.~\ref{fig;dressed-t-propagator} are the diagrams for the 
dressed hypertriton ($t$ field) propagator as a $\Lambda d$ composite state. 
Then the renormalized dressed hypertriton propagator is obtained as
\begin{widetext}
\begin{eqnarray}
D_t (p_0^{},\bm{p}) =
\frac{2\pi}{\mu_{\Lambda d}^{} y_t^2}
\left\{
\frac{1}{a_{\Lambda d}^{}} + \frac12 r_{\Lambda d}^{} \left[
-2\mu_{\Lambda d}^{} \left( p_0^{} - \frac{1}{2(m_\Lambda^{} +m_d^{})} \bm{p}^2
+ i \epsilon \right) \right]
- \sqrt{ -2\mu_{\Lambda d}^{} \left( p_0^{} - \frac{1}{2(m_\Lambda^{} + m_d^{})} \bm{p}^2
+ i\epsilon \right) } 
\right\}^{-1}
\label{eq;Dt}
\end{eqnarray}
\end{widetext}
with
\begin{eqnarray}
y_t = - \frac{1}{\mu_{\Lambda d}^{} }
\sqrt{ \frac{2\pi}{r_{\Lambda d}^{}} }\,,
\end{eqnarray}
where $\mu_{\Lambda d}^{}$ is the reduced mass of the $\Lambda d$ system, i.e.,
$\mu_{\Lambda d}^{} = m_\Lambda^{} m_d^{}/(m_\Lambda^{} + m_d^{})$,
and $a_{\Lambda d}^{}$ and $r_{\Lambda d}^{}$ are the effective range parameters
of the $S$-wave $\Lambda$-$d$ scattering in the hypertriton channel.
In Ref.~\cite{Hammer01}, these effective range parameters are estimated 
as $a_{\Lambda d}^{} = 16.8 \substack{+ 4.4 \\ - 2.4}$~fm, and 
$r_{\Lambda d}^{} = 2.3 \pm 0.3$~fm, which leads to 
$\gamma_{\Lambda d}^{} = 1/a_{\Lambda d}^{} +r_{\Lambda d}^{} \gamma_{\Lambda d}^2/2 
\simeq 12.8~\mbox{\rm MeV}$ when we use the central values of the parameters.
This value is consistent with the one given in Sec.~\ref{sec:Introduction} 
within error.

Since there exists a bound state for hypertriton, the propagator should have a pole at 
$k=i\gamma_{\Lambda d}^{}$ and we may rewrite the on-energy-shell dressed propagator as
\begin{eqnarray}
D_t(E) &=& \frac{2\pi}{\mu_{\Lambda d}^{} y_t^2}
\left[
 \gamma_{\Lambda d}^{} - \frac12 r_{\Lambda d}^{} 
\left( k^2 + \gamma_{\Lambda d}^2 \right) + i k
\right]^{-1},
\end{eqnarray}
where $E = k^2/(2\mu_{\Lambda d}^{})$.
Furthermore, near the pole, the propagator can be further simplified as 
\begin{eqnarray}
D_t (E) \simeq \frac{Z_{\Lambda d}}{E+B_{\Lambda}}\,
\quad \mbox{\rm with} \quad
Z_{\Lambda d} = \frac{\gamma_{\Lambda d}^{} r_{\Lambda d}^{}}
{1 - \gamma_{\Lambda d}^{} r_{\Lambda d}^{} } ,
\end{eqnarray}
where $Z_{\Lambda d}$ is the wave function normalization factor
of the hypertriton as a $\Lambda d$ system.
Since the inverse of the effective range has a large scale, $r_{\Lambda d}^{-1} \simeq 86$~MeV, 
one can see that the KSW counting rules, where the propagator and $Z_{\Lambda d}$ 
are expanded in terms of $r_{\Lambda d}^{}$, would be a good approximation, which can
be seen from the fact that $\gamma_{\Lambda d}^{} r_{\Lambda d}^{} \simeq 0.16 < 1/3$.


\section{Three-body part}
\label{sec:3body}

\begin{figure}[t]
\begin{center}
\includegraphics[width=0.9\columnwidth]{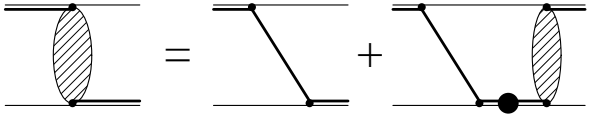}
\caption{
Diagrams of the integral equation for $S$-wave scattering of hypertriton and $\Lambda$ 
for spin singlet ($S=0$) channel.
See the caption of Fig.~\ref{fig;dressed-t-propagator} as well.
}
\label{fig;singlet-IE}
\end{center}
\end{figure}

In this Section, we construct the integral equations for $S$-wave scattering of hypertriton and 
$\Lambda$, which has two spin channels, $S=0$ and 1, because both the hypertriton 
and $\Lambda$ have spin-1/2.
For $S=0$ channel, the amplitude $t(p,k;E)$ consists of hypertriton-$\Lambda$ 
channel only.
In Fig.~\ref{fig;singlet-IE}, 
diagrams of the integral equation for the scattering amplitude are
shown, which lead to
\begin{widetext}
\begin{eqnarray}
t(p,k;E) =
-3K_{(a)}(p,k;E)
+ \frac{1}{2\pi^2} \int^{\Lambda_c}_0 d \ell \, \ell^2 \, 3K_{(a)}(p,\ell;E) 
D_t\left( E-\frac{\ell^2}{2m_\Lambda},\bm{\ell}\right) t(\ell,k;E) 
\end{eqnarray} 
\end{widetext}
with the one-deuteron-exchange interaction $K_{(a)}(p,l;E)$, 
\begin{eqnarray}
K_{(a)}(p,\ell;E) &=& \frac13\frac{m_d^{} y_t^2}{2p \ell} \ln\left(
\frac{\frac{m_d^{}}{2\mu^{}_{\Lambda d}}(p^2+\ell^2) + p \ell - m_d^{} E}
     {\frac{m_d^{}}{2\mu_{\Lambda d}^{}}(p^2+\ell^2) - p \ell - m_d^{} E}
\right) ,
\nonumber \\
\end{eqnarray}
where $p$ and $k$ are relative off-shell and on-shell momenta of 
hypertriton-$\Lambda$ scattering in the CM frame, respectively, 
and $E$ is the total energy, 
\begin{eqnarray}
E &=& - \frac{\gamma_{\Lambda d}^2}{2\mu_{\Lambda d}^{} }
+ \frac{1}{2 \mu_{\Lambda(\Lambda d)^{} }} k^2 ,
\end{eqnarray}
with $\mu_{\Lambda(\Lambda d)}^{}$ being the reduced mass of the 
$\Lambda$-$(\Lambda d)$ system so that $\mu_{\Lambda(\Lambda d)}^{} = 
m_\Lambda^{} (m_\Lambda^{} + m_d^{})/(2m_\Lambda^{} + m_d^{} )$. 
A sharp cutoff momentum $\Lambda_c$ was introduced as before in the integral equation. 
However, as we shall see below, the integral equation is insensitive to the value 
of $\Lambda_c$, 
which weakens the necessity of three-body contact interactions.

\begin{figure*}[t]
\begin{center}
\includegraphics[width=0.80\textwidth]{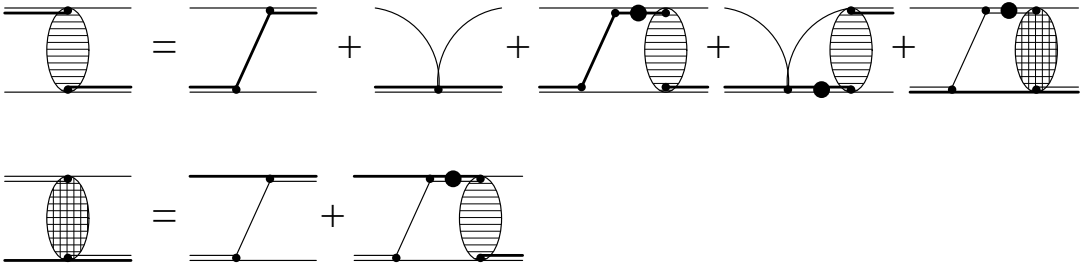}
\caption{
Diagrams of coupled integral equations for 
$S$-wave scattering of hypertriton and $\Lambda$ 
for spin triplet ($S=1$) channel.
See the captions of Figs.~\ref{fig;dressed-propagator}
and \ref{fig;dressed-t-propagator} as well.
}
\label{fig;triplet-integral-equations}
\end{center}
\end{figure*}

For $S=1$ channel, however, we have two scattering amplitudes, namely,
$a(p,k;E)$ for the spin triplet $\Lambda t$ ($\Lambda$ and hypertriton) 
cluster channel and $b(p,k;E)$ that connects the $\Lambda t$ 
cluster channel to the $d s$ (deuteron and the $\Lambda\Lambda$ dibaryon)  
cluster channel.
In Fig.~\ref{fig;triplet-integral-equations}, diagrams
of coupled integral equations are presented, from which we obtain
\begin{widetext}
\begin{eqnarray}
a(p,k;E) &=& 
K_{(a)}(p,k;E)  - \frac{g_1(\Lambda_c)}{\Lambda_c^2}
- \frac{1}{2\pi^2} \int^{\Lambda_c}_0 d\ell \, \ell^2
\left[K_{(a)}(p,\ell;E) -\frac{g_1^{}(\Lambda_c)}{\Lambda_c^2} \right]
D_t\left( \textstyle E -\frac{\ell^2}{2m_\Lambda}^{} , \bm{\ell} \right) a(\ell,k;E)
\nonumber \\ && \mbox{}
- \frac{1}{2\pi^2}\int^{\Lambda_c}_0 d\ell \, \ell^2
K_{(b1)}(p,\ell;E) D_s\left( \textstyle E -\frac{\ell^2}{2m_d}, \bm{\ell} \right) b(\ell,k;E) ,
\nonumber \\
b(p,k;E) &=& K_{(b2)}(p,k;E) 
- \frac{1}{2\pi^2} \int^{\Lambda_c}_0 d\ell \, \ell^2
K_{(b2)}(p,\ell;E) D_t\left( \textstyle E -\frac{\ell^2}{2m_\Lambda}^{} , \bm{\ell} \right) 
a(\ell,k;E) ,
\label{eq:integral}
\end{eqnarray}
with one-$\Lambda$-exchange interactions $K_{(b1)}(p,\ell;E)$ and $K_{(b2)}(p,\ell;E)$,
which read
\begin{eqnarray}
K_{(b1)}(p,\ell;E) &=& - \sqrt{\frac{2}{3}} \frac{m_\Lambda^{} y_s^{} y_t^{}}{2p\ell}
\ln\left[ \frac{p^2 + \frac{m_\Lambda^{} }{2\mu_{d\Lambda}^{}} \ell^2 + p \ell - m_\Lambda^{} E}
   {p^2 + \frac{m_\Lambda^{} }{2\mu_{d\Lambda}^{}} \ell^2 - p \ell - m_\Lambda^{} E}
\right] ,
\\
K_{(b2)}(p,\ell;E) &=& - \sqrt\frac{2}{3} \frac{m_\Lambda^{} y_s^{} y_t^{}}{2p \ell}
\ln\left[
\frac{\frac{m_\Lambda^{} }{2\mu_{d\Lambda}^{} } p^2 + \ell^2 + p \ell - m_\Lambda^{} E}
  {\frac{m_\Lambda^{}}{2\mu_{d\Lambda}^{}} p^2 + \ell^2 - p \ell - m_\Lambda^{} E}
\right] .
\end{eqnarray}
\end{widetext}
In Eq.~(\ref{eq:integral}), we have introduced the three-body contact interaction that contains
the coupling constant $g_1^{}(\Lambda_c)$.%
\footnote{The coupling constant $g_1^{} (\Lambda_c)$ is a dimensionless quantity. 
}
As we shall see below, the integral equations depend on the cutoff $\Lambda_c$ and 
$g_1^{} (\Lambda_c)$ accounts for the high momentum effects above $\Lambda_c$.


\section{Numerical results}
\label{sec:results}

\subsection{\boldmath $S$-wave scattering of hypertriton and 
$\Lambda$ in $S=0$ channel}

In the dressed hypertriton propagator $D_t$ given in Eq.~(\ref{eq;Dt}), 
there are two singularities at $\ell \simeq 13$~MeV and $\ell \simeq 172$~MeV 
when $E=0$ in Eq.~(\ref{eq:integral}). 
The first one corresponds to the binding momentum of the hypertriton in the 
$\Lambda$-$d$ system and the second one to an unphysical deeply-bound state. 
We avoid the effect from the unphysical deeply bound state 
by expanding the effective range correction, as mentioned above,
employing the KSW counting rules.

\begin{figure}[t]
\begin{center}
\includegraphics[width=1.0\columnwidth]{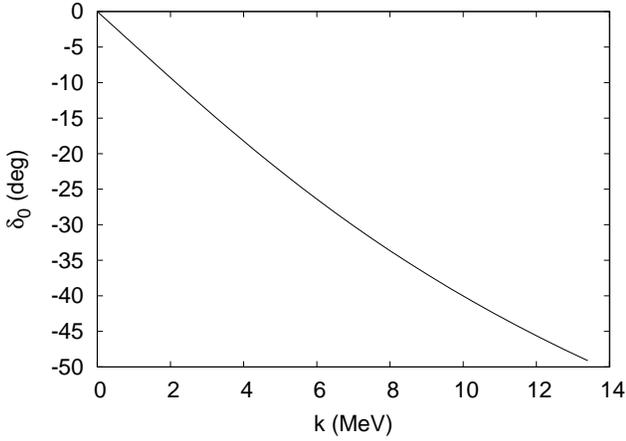}
\caption{ 
Phase shift $\delta_0$ (in degrees) of the $S$-wave hypertriton-$\Lambda$
scattering in the spin singlet channel as a function of momentum $k$ (in MeV).
}
\label{fig;del0}
\end{center}
\end{figure}

The on-shell scattering matrix is given by
\begin{eqnarray}
T(k,k) = \sqrt{Z_{\Lambda d}^{}}\, t(k,k;E) \sqrt{Z_{\Lambda d}} \, ,
\end{eqnarray}
and thus the integral equation in terms of the half-off-shell scattering matrix 
at LO reads
\begin{widetext}
\begin{eqnarray}
T(p,k) &=& 
-3 \, \gamma_{\Lambda d}^{} r_{\Lambda d}^{} K_{(a)}(p,k;E)
\nonumber \\ && \mbox{}
-\frac{3}{2\pi^2} \mu_{\Lambda(\Lambda d)}^{} r_{\Lambda d}^{}
\int^{\Lambda_c}_0 d \ell\, K_{(a)}(p,\ell;E)
\left\{ 
\gamma_{\Lambda d}^{}
+ \sqrt{\gamma_{\Lambda d}^2 + \frac{\mu_{\Lambda d}^{} }{\mu_{\Lambda(\Lambda d)}^{} }
\left( \ell^2-k^2 \right) }
\right\}
\frac{\ell^2 \, T(\ell,k)}{\ell^2-k^2-i\epsilon}  \, .
\end{eqnarray}
\end{widetext}
This shows that the integral equation is expressed in terms of two parameters, namely,
$\gamma_{\Lambda d}^{}$ and $\Lambda_c$, in addition to the deuteron and $\Lambda$ 
masses. 
As mentioned before, this integral equation is insensitive to the value of $\Lambda_c$ 
and thus the scattering in the $S=0$ channel is well controlled by 
one effective range parameter, $\gamma_{\Lambda d}^{}$.

The scattering length $a_0^{}$ of the $S$-wave hypertriton-$\Lambda$ scattering 
in the $S=0$ channel is then computed by taking the limit for the on-shell momentum 
$k \to 0$, which leads to 
$T(0,0) = - \frac{2\pi}{\mu_{\Lambda(\Lambda d)}^{}} a_0^{}$.  
Here, we introduce the half-off-shell scattering length $a(p,0)$ as
\begin{eqnarray}
a(p,0) = - \frac{\mu_{\Lambda(\Lambda d)}^{}}{2\pi}\, T(p,0) ,
\end{eqnarray}
so that it reduces to the scattering length as $a_0^{} = a(0,0)$.

We numerically calculate the off-diagonal part of the scattering length 
$a_0^{}(p,0)$ with $\Lambda_c \sim 170$~MeV to find that the off-diagonal 
part of the scattering length becomes indeed very small when the off-shell momentum $p$ 
is larger than the large scale $\Lambda_H \sim \gamma \simeq 45.7$~MeV.
We also calculate the scattering length $a_0^{} (0,0)$ as a function of the cutoff 
$\Lambda_c$ to find that $a (0,0)$ is nearly independent of the cutoff 
if it is relatively small such as $\Lambda_c \simeq 20$~MeV.
Therefore, the $S$-wave hypertriton-$\Lambda$ scattering 
in spin singlet channel would be well described by considering 
the cutoff region of $\Lambda_c \simeq \Lambda_H$. 
From this procedure we obtain
\begin{eqnarray}
a_0^{} = 16.0 \pm 3.0 \mbox{ fm} ,
\end{eqnarray} 
which is our prediction on the scattering length, where the error was estimated 
from the uncertainties in $\gamma_{\Lambda d}^{}$.%
\footnote{
Alternatively, one may include the effective range $r_{\Lambda d}^{}$ in the 
dressed propagator, as  in Refs.~\cite{BV98a,Ando13} for the studies on the 
$S$-wave neutron-deuteron scattering in spin quartet channel within pionless EFT.
If we take this procedure, we would obtain
$a_0^{} = 17.3 \pm 2.9$~fm.
}

In Fig.~\ref{fig;del0}, the calculated phase shift $\delta_0$ of the $S$-wave 
hypertriton-$\Lambda$ scattering in the spin singlet channel is presented as 
a function of $k$.
The form of the calculated phase shift $\delta_0$ determines the 
two effective range parameters as 
$a_0^{} \simeq 16.0$~fm and $r_0^{} \simeq 2$~fm. 
In addition, we find no limit-cycle in the numerical calculation of the integral 
equation within the range up to $\Lambda_c \sim  10^{8}$~MeV.

\begin{widetext}
\subsection{\boldmath $_{\Lambda\Lambda}^{\ \ 4}$H bound state and $S$-wave 
scattering of hypertriton-$\Lambda$ in $S=1$ channel}

For the spin triplet channel, the coupled integral equations can be rewritten 
in terms of the half-off-shell scattering amplitudes $a_1^{} (p,k)$ and 
$b_1^{} (p,k)$ which are defined by
\begin{eqnarray}
a_1^{} (p,k) &=& - \frac{Z_{\Lambda d}}{2\pi} \mu_{\Lambda (\Lambda d)}^{}
\left[ K_{(a)}(p,k;E) - \frac{g_1^{}(\Lambda_c)}{\Lambda_c^2} \right]
- \frac{1}{2\pi^2} \int_0^{\Lambda_c} d \ell \, \ell^2 \left[ K_{(a)}(p,\ell;E) 
-\frac{g_1^{} (\Lambda_c)}{\Lambda_c^2} \right] 
D_t\left( \textstyle E - \frac{\ell^2}{2m_\Lambda^{}},\bm{\ell}\right) 
a_1^{} (\ell,k)
\nonumber \\ && \mbox{}
- \frac{1}{2\pi^2} \int_0^{\Lambda_c} d\ell \, \ell^2 K_{(b1)}(p,\ell;E) 
D_s\left( \textstyle E - \frac{\ell^2}{2m_d^{}},\bm{\ell} \right) b_1^{} (\ell,k) ,
\\ 
b_1(p,k) &=& - \frac{Z_{\Lambda d}}{2\pi} \mu_{\Lambda (\Lambda d)}^{} K_{(b2)}(p,k;E) 
- \frac{1}{2\pi^2} \int_0^{\Lambda_c} d\ell \, \ell^2 K_{(b2)}(p,\ell;E) 
D_t\left( \textstyle E - \frac{\ell^2}{2m_\Lambda^{}},\bm{\ell} \right) a_1^{} (\ell,k) ,
\end{eqnarray}
\end{widetext}
with the normalizations 
\begin{eqnarray}
a_1^{} (k,k) &=& \sqrt{Z_{\Lambda d}}\,a(k,k) \sqrt{Z_{\Lambda d}}\,,
\nonumber \\
b_1^{}(k,k) &=& \sqrt{Z_{\Lambda d}}\,b(k,k) \sqrt{Z_{\Lambda d}}\,.
\end{eqnarray}
The scattering length $a_1^{}$ is then defined as
\begin{equation}
a_1^{} = -\frac{\mu_{\Lambda(\Lambda d)^{}} }{2\pi} a_1^{} (0,0).
\end{equation}

Because the effect from the unphysical singularities in the dressed dibaryon and 
hypertriton propagators ($D_s$ and $D_t$) to the scattering length $a_1^{}$
is significant, we employ the KSW counting rules and expand the propagators 
and the wave function normalization factor $Z_{\Lambda d}$ in terms of the 
effective ranges $r_{\Lambda d}^{}$ and $r_{\Lambda\Lambda}^{}$,
as discussed in Sec.~\ref{sec:Introduction}.
Therefore, at LO, the propagators $D_t$ and $D_s$ and the wave function normalization 
factor $Z_{\Lambda d}$ are written as
\begin{widetext}
\begin{eqnarray}
D_t^{\rm LO} \left( \textstyle E -\frac{\ell^2}{2m_\Lambda^{} },\bm{\ell} \right) 
&=&
- \frac{2\pi \, \mu_{\Lambda(\Lambda d)}^{} }{\mu_{\Lambda d}^2 y_t^2}
\left[ \gamma_{\Lambda d}^{} + \sqrt{\gamma_{\Lambda d}^2 
+\frac{\mu_{\Lambda d}^{} }{ \mu_{\Lambda(\Lambda d)}^{}} (\ell^2-k^2)}
\right] \frac{1}{\ell^2-k^2 -i\epsilon}\,,
\\
D_s^{\rm LO} \left( \textstyle E - \frac{\ell^2}{2m_d^{}},\bm{\ell} \right) &=& 
\frac{4\pi}{m_\Lambda^{} y_s^2} 
\left[
\frac{1}{a_{\Lambda\Lambda}^{} }
- \sqrt{ \frac{m_\Lambda^{} }{2\mu_{\Lambda d}^{}} \gamma_{\Lambda d}^2
- \frac{m_\Lambda^{}}{2}\left(
\frac{\ell^2}{\mu_{d(\Lambda\Lambda)}^{} }
-\frac{k^2}{\mu_{\Lambda(\Lambda d)}^{}} \right) }
\right]^{-1} ,
\\
Z_{\Lambda d}^{\rm LO} &=& \gamma_{\Lambda d}^{} \, r_{\Lambda d}^{} \,,
\end{eqnarray}
where $\mu_{d(\Lambda\Lambda)}^{}$ is the reduced mass of the 
$d$-$(\Lambda\Lambda)$ system,
$\mu_{d(\Lambda\Lambda)}^{} = 2m_\Lambda^{} m_d^{}/(2m_\Lambda^{} + m_d^{})$.
\end{widetext}

In addition to the masses,  therefore,  we have four parameters, namely, 
$\gamma_{\Lambda d}^{}$, $a_{\Lambda\Lambda}^{}$, $g_1^{} (\Lambda_c)$, 
and $\Lambda_c$.%
\footnote{
In principle, the integral equation depends on the effective ranges $r_{\Lambda d}^{}$ and 
$r_{\Lambda\Lambda}^{}$ through the coupling constants $y_{t,s}$ and the normalization 
factor $Z_{\Lambda d}^{\rm LO}$. 
But this dependence is canceled or included in the normalization of the amplitude $b_1$,
and, therefore, they do not appear in the final expressions. 
}
In the present work, we fix $\gamma_{\Lambda d}^{}$ by the hypertriton binding energy.
The parameter $a_{\Lambda\Lambda}^{}$ may be determined from other available
empirical information.
However, there exists no available information from the three-body system to 
constrain the value of $g_1^{} (\Lambda_c)$.
In the present work, therefore, instead of studying the energy levels of 
the $_{\Lambda\Lambda}^{\ \, 4}$H hypernucleus,
we examine the effect of the coupling $g_1^{} (\Lambda_c)$ in this system.

\subsubsection{Scattering length $a_1^{}$ without three-body contact interaction}

We first consider the case when $g_1^{}(\Lambda_c) = 0$
and calculate the two-$\Lambda$ separation energy $B_{\Lambda\Lambda}$ in the 
${}_{\Lambda\Lambda}^{\ \, 4}$H bound state and the scattering length $a_1^{}$ of 
the $S$-wave hypertriton-$\Lambda$ scattering for the spin triplet channel at LO.
In this case we find that there is no bound state formed with the cutoff value 
in the range of $\Lambda_c = 50 \sim 300$~MeV.


\begin{figure}[t]
\begin{center}
\includegraphics[width=1.0\columnwidth]{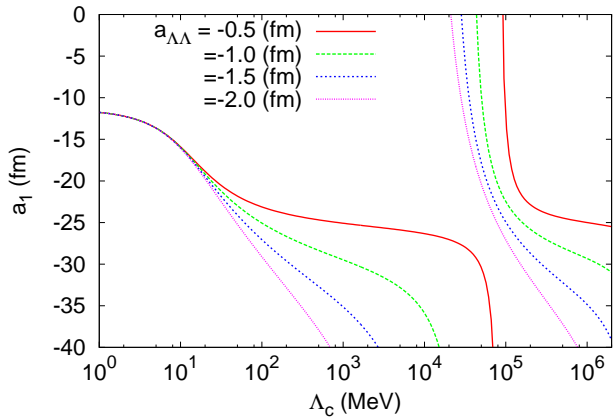}
\caption{
(Color online)
Scattering length $a_1^{}$ of $S$-wave hypertriton-$\Lambda$ scattering in the $S=1$ 
channel at leading order as a function of $\Lambda_c$ for 
$a_{\Lambda\Lambda}^{} =-0.5, -1.0, -1.5,-2.0$~fm.}
\label{fig;a1dvsLam-LO}
\end{center}
\end{figure}

In Fig.~\ref{fig;a1dvsLam-LO}, we present our results for the LO scattering 
length $a_1^{}$ with several values of $a_{\Lambda\Lambda}^{}$, namely, 
$a_{\Lambda\Lambda}^{} =-0.5$, $-1.0$, $-1.5$, $-2.0$~fm,
as a function of the momentum cutoff $\Lambda_c$.
This shows that the calculated $a_1^{}$ curves show a significant dependence on 
$\Lambda_c$ as well as on $a_{\Lambda\Lambda}^{}$.
The $a_{\Lambda\Lambda}^{}$-dependence of $a_1^{}$ becomes more significant 
when $\Lambda_c$ is larger than $\Lambda_H$ as shown in Fig.~\ref{fig;a1dvsLam-LO}. 
When the cutoff parameter $\Lambda_c$ is down close to the large scale of 
the theory, i.e., $\Lambda_c \simeq \Lambda_H \sim 45.7$~MeV, 
such a dependence becomes mild.
We then obtain negative values for the scattering length, namely, 
$a_1^{} \simeq -21.7$, $-22.7$, $-23.8$, $-24.8$~fm for 
$a_{\Lambda\Lambda}^{} = -0.5$, $-1.0$, $-1.5$, $-2.0$~fm,
respectively, with $\Lambda_c = 45.7$~MeV.
Since $a_1^{}$ is negative and its magnitude is large,
it may imply a formation of a quasi-bound state.

As $\Lambda_c$ increases, $a_1^{}$ decreases until it shows a pole-structure at
around $\Lambda_c \sim 80$, $33$, $17$, $10$~GeV depending on the value of 
$a_{\Lambda\Lambda}^{}$.
After passing the pole, $a_1^{}$ changes the sign as shown in Fig.~\ref{fig;a1dvsLam-LO}.
This corresponds to a formation of a bound state with zero binding energy at such 
a huge cutoff.
In other words, the one-deuteron-exchange interaction has a sensitivity to $\Lambda_c$ 
and it becomes attractive enough to make a bound state at the asymptotic limit of 
the cutoff.

To make the result cutoff-independent, however, one needs to promote the three-body 
contact interaction at LO so that the cutoff dependence is controlled by the additional 
coupling constant~\cite{BHV99}. 
We work on in this scheme below.

\subsubsection{$_{\Lambda\Lambda}^{\ \ 4}$H bound state
with three-body contact interaction}

\begin{figure}[t]
\begin{center}
\includegraphics[width=1.0\columnwidth]{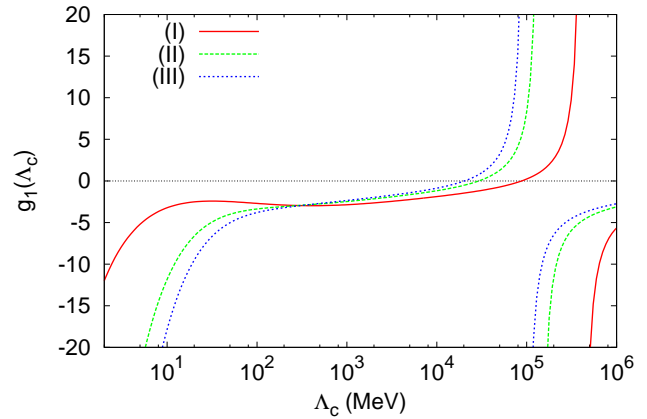}
\caption{
(Color online)
Coupling $g_1^{}(\Lambda_c)$ of three-body contact interaction as a function of the 
cutoff $\Lambda_c$ which produces a bound state of $_{\Lambda\Lambda}^{\ \ 4}$H  
with three different sets of $B_{\Lambda\Lambda}$ and $a_{\Lambda\Lambda}^{}$. 
See the text for the parameter sets (I), (II), and (III).
}
\label{fig;gvsLam}
\end{center}
\end{figure}

We now consider the case with $g_1^{}(\Lambda_c) \neq 0$ to investigate its role
in the ${}_{\Lambda\Lambda}^{\ \, 4}$H hypernucleus.
Since there is no experimental information to constrain the value of 
$g_1^{}(\Lambda_c)$, we adopt the values of this coupling constant
determined as follows.
We first assume a formation of the $_{\Lambda\Lambda}^{\ \ 4}$H bound state due to the 
three-body-contact interaction and fit $g_1^{}(\Lambda_c)$ to reproduce the potential model 
results of Refs.~\cite{FG02,NAM02}.
To be specific, we choose the following three sets for 
$B_{\Lambda\Lambda}$ and $a_{\Lambda\Lambda}^{}$:
\begin{eqnarray}
& \mbox{(I) } & B_{\Lambda\Lambda}^{} \simeq 0.2 \mbox{ MeV and }
a_{\Lambda\Lambda}^{} = -0.5 \mbox{~fm},
\nonumber \\ &
\mbox{(II) } & B_{\Lambda\Lambda} \simeq 0.6 \mbox{~MeV and }
a_{\Lambda\Lambda}^{} = -1.5 \mbox{~fm},
\nonumber \\ &
\mbox{(III) } & B_{\Lambda\Lambda} \simeq 1.0 \mbox{~MeV and }
a_{\Lambda\Lambda}^{} = -2.5 \mbox{~fm}.
\label{eq:param}
\end{eqnarray}

In Fig.~\ref{fig;gvsLam}, we show the calculated strength of the three-body contact 
interaction $g_1^{}(\Lambda_c)$ as a function of $\Lambda_c$, 
which can reproduce the three parameter sets of Eq.~(\ref{eq:param}).
One can see that the curves of $g_1^{} (\Lambda_c)$ are rather mildly varying at 
$\Lambda_c = 10 \sim 10^4$~MeV, and each curve has a singularity at  
$\Lambda_c \sim 10^5$~MeV indicating the possibility of the first cycle of the limit-cycle.
This implies that the one-deuteron-exchange interaction for the $S=1$ channel contains 
an attractive (singular) interaction at very high momentum, say, $\Lambda_c \sim 10^5$~MeV.
This property has also been observed in the calculation of $a_1^{}$ as shown 
in Fig.~\ref{fig;a1dvsLam-LO}.
At such a very high momentum, however, the applicability of the present theory, a very low 
energy EFT, cannot be guaranteed and thus the mechanisms of the formation of 
a bound state must have different origins.
We note, on the other hand, that, if we choose $g_1^{} (\Lambda_c) \simeq -2$ 
or smaller at $\Lambda_c \sim 50$~MeV in the coupled integral equations,
a bound state can be created.
Such a value of $g_1^{} (\Lambda_c)$ is in a natural size and may be generated 
from the mechanisms of high energy such as $\sigma$-meson exchange or 
two-pion exchange near the intermediate range of nuclear force, i.e., 
$\Lambda_c =300 \sim 600$~MeV.


\begin{figure}[t]
\begin{center}
\includegraphics[width=1.0\columnwidth]{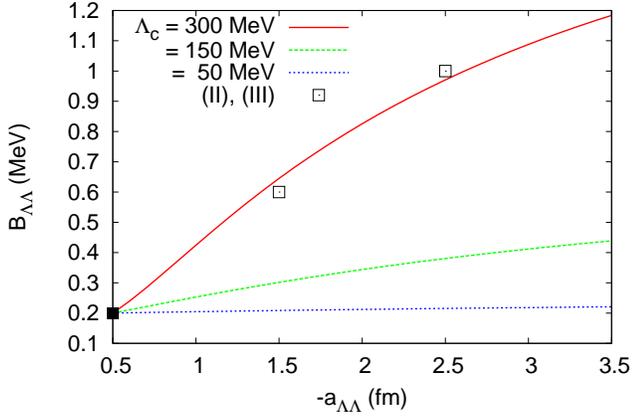}
\caption{ 
(Color online)
Calculated two-$\Lambda$ separation energy $B_{\Lambda\Lambda}$ from 
${}_{\Lambda\Lambda}^{\ \ 4}$H bound state as a function of the scattering length 
$a_{\Lambda\Lambda}^{}$ of the $S$-wave $\Lambda\Lambda$ scattering for the ${}^1S_0$ 
channel with the cutoff values $\Lambda_c=50$, 150, 300~MeV.
The value of $g_1^{} (\Lambda_c)$ of all three curves is fitted 
at the point (I): 
$B_{\Lambda\Lambda}=0.2$~MeV and $a_{\Lambda\Lambda}^{} = -0.5$~fm,
marked by a filled square. 
The points (II) and (III) are also included as blank squares in the figure. 
}
\label{fig;BLLvsaLL}
\end{center}
\end{figure}

In order to study the correlation between $B_{\Lambda\Lambda}$ and 
$a_{\Lambda\Lambda}^{}$, we calculate $B_{\Lambda\Lambda}$ as a function of 
$a_{\Lambda\Lambda}^{}$ and show the results
in Fig.~\ref{fig;BLLvsaLL} for various cutoff values, i.e., 
$\Lambda_c=50$, $150$, $300$~MeV. 
Here, the coupling $g_1^{} (\Lambda_c)$ is fixed by using the parameter set (I), i.e.,
$B_{\Lambda\Lambda}=0.2$~MeV and $a_{\Lambda\Lambda}=-0.5$~fm, which is
marked by a filled square in Fig.~\ref{fig;BLLvsaLL}.
This is achieved with $g_1^{} (\Lambda_c) \simeq -2.48$, $-2.83$, $-2.96$ 
for $\Lambda_c=50$, $150$, $300$~MeV, respectively.
Once the starting values are fixed, we vary the value of $a_{\Lambda\Lambda}^{}$ 
for a fixed value of $\Lambda_c$, which changes the values of $B_{\Lambda\Lambda}$.
We then find that the behaviors of the $B_{\Lambda\Lambda}$ curves 
as functions of $a_{\Lambda\Lambda}$ are quite sensitive to the values of 
the cutoff $\Lambda_c$.  
For example, when we choose $\Lambda_c \simeq \Lambda_H$, 
i.e., $\Lambda_c=50$~MeV, $B_{\Lambda\Lambda}$ is insensitive to the value of
$a_{\Lambda\Lambda}^{}$ and makes a nearly flat curve as shown by the dotted line
in Fig.~\ref{fig;BLLvsaLL}.
However, with a larger cutoff value, $\Lambda_c = 300$~MeV, $B_{\Lambda\Lambda}$
strongly depends on $a_{\Lambda\Lambda}^{}$ and we can fairly well 
reproduce the $a_{\Lambda\Lambda}^{}$-dependence of $B_{\Lambda\Lambda}$ 
obtained by Filikhin and Gal~\cite{FG02} or Nemura {\it et al.}~\cite{NAM02}.

This may imply that the main part of the correlation between $B_{\Lambda\Lambda}$ 
and $a_{\Lambda\Lambda}$ in potential model calculations is related to the high 
momentum part and, when we choose the cutoff $\Lambda_c \simeq \Lambda_H$, 
the mechanisms with high momentum are integrated out and their effects are
absorbed by the renormalized three-body contact interaction $g_1^{}(\Lambda_c)$.
Thus we do not have the dynamics that is sensitive to the high momentum regime
and this leads to the cutoff-insensitive results. 
Therefore, when we choose $\Lambda_c\simeq\Lambda_H=50$~MeV in our cluster EFT, 
the theory does not to adequately probe the $\Lambda$-$\Lambda$ interactions,
but, when we choose $\Lambda_c = 300$~MeV, we can 
fairly well reproduce the results obtained in the potential model calculations.
However, in the latter case, the theory becomes inconsistent because of neglecting other 
mechanisms relevant in the high momentum region, such as the channels of deuteron
break-up into two nucleons and of meson-exchanges among baryons.


\begin{figure}[t]
\begin{center}
\includegraphics[width=1.0\columnwidth]{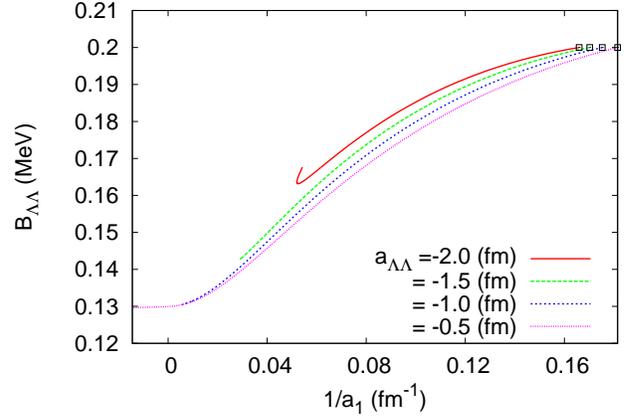}
\caption{ 
(Color online)
Correlations between $B_{\Lambda\Lambda}$ and $1/a_1^{}$ with 
$a_{\Lambda\Lambda}^{} = -2.0$, $-1.5$, $-1.0$, $-0.5$~fm.
The coupling $g_1^{} (\Lambda_c)$ is fixed by $B_{\Lambda\Lambda}=0.2$~MeV and 
$\Lambda_c=50$~MeV, marked by open squares in the upper-right corner, 
for each value of $a_{\Lambda\Lambda}^{}$.
The curves are obtained by varying $\Lambda_c$ from 50~MeV to 300~MeV.
}
\label{fig;BLLvsa1d-aLLs}
\end{center}
\end{figure}

In Fig.~\ref{fig;BLLvsa1d-aLLs}, we present our results on the correlation between 
$B_{\Lambda\Lambda}$ and $1/a_1^{}$ with four values of $a_{\Lambda\Lambda}^{}$
where $g_1^{} (\Lambda_c)$ is fixed by using the condition that $B_{\Lambda\Lambda}=0.2$~MeV 
at $\Lambda_c=50$~MeV. 
Thus with $\Lambda_c=50$~MeV, we have $g_1^{} = -2.48$, $-2.45$, $-2.43$, $-2.40$ 
for $a_{\Lambda\Lambda}^{} =-0.5$, $-1.0$, $-1.5$, $-2.0$~fm, respectively.
Then the curves are obtained by varying $\Lambda_c$ from 50~MeV to 300~MeV
with the fixed values of $g_1^{}$ determined at $\Lambda_c=50$~MeV.
We find that, at $\Lambda_c = 50$~MeV, which gives the starting points of the curves 
at the top right corner (marked by open squares), the calculated scattering 
length $a_1^{}$  at LO turned out to be positive due to the existence of
${}_{\Lambda\Lambda}^{\ \,4}$H bound state and the positions of these points are 
not sensitive to the value of $a_{\Lambda\Lambda}^{}$, as was seen in 
Fig.~\ref{fig;BLLvsaLL} for the case of $B_{\Lambda\Lambda}$ with $\Lambda_c=50$~MeV.
Thus we have $a_{1^{} }\sim 5.7$~fm corresponding to 
$B_{\Lambda\Lambda}\simeq 0.2$~MeV. 
By increasing the cutoff values, we obtain the lower values of $B_{\Lambda\Lambda}$.
When $a_{\Lambda\Lambda}^{} =-0.5$~fm, the $_{\Lambda\Lambda}^{\ \, 4}$H bound state 
eventually becomes unbound, and when $a_{\Lambda\Lambda}^{} =-2.0$~fm, 
$B_{\Lambda\Lambda}$ has a minimum and then starts to increase with increasing cutoff.
We also find that the correlations do not show the sensitivity to $a_{\Lambda\Lambda}^{}$.


\section{Summary and Discussion}
\label{sec:discussion}

In the present work, we studied the ${}_{\Lambda\Lambda}^{\ \ 4}$H bound state
and $S$-wave hypertriton-$\Lambda$ scattering for spin singlet and triplet channels 
below the hypertriton breakup momentum in Halo EFT at LO by treating the 
${}_{\Lambda\Lambda}^{\ \ 4}$H system 
as a three-body $\Lambda \Lambda d$ cluster system.
In this approach, the hypertriton breakup momentum 
$\gamma_{\Lambda d}^{} \simeq 13.4$~MeV is chosen to be the typical scale $Q$
of the theory, 
whereas the deuteron binding momentum $\gamma \simeq 45.7$~MeV to be 
the high momentum scale $\Lambda_H$.
Thus, in such a small typical momentum scale, the deuteron is not broken into 
two nucleons, which justifies the treatment of the deuteron field as 
a cluster (elementary) field.
Furthermore, our expansion parameter is $Q/\Lambda_H \sim
\gamma_{\Lambda d}^{}/\gamma \sim 1/3$.

For the spin singlet channel of the $S$-wave hypertriton-$\Lambda$ scattering, 
the amplitude is nearly independent of the cutoff, 
thus there is no need to introduce the three-body contact interaction at LO. 
Consequently, the integral equation at LO is well described by one effective range 
parameter, $\gamma_{\Lambda d}^{}$.
This leads to the value of the scattering length $a_0^{}$ of the $S$-wave 
hypertriton-$\Lambda$ scattering for the spin singlet channel as 
$a_0^{} = 16.0 \pm 3.0$~fm.
We also found no bound state in this channel at LO.

For the spin triplet channel of the $S$-wave scattering of 
hypertriton and $\Lambda$, the scattering amplitudes are obtained through two 
coupled integral equations.
We find that when the cutoff parameter $\Lambda_c$ is close to the asymptotic 
limit, the coupling of the three-body contact interaction, i.e.,  
$g_1^{}(\Lambda_c)$, exhibits the limit-cycle, and thus the three-body contact interaction 
should be included in the spin triplet channel. 
Consequently the coupled integral equations are represented by four parameters, 
$\gamma_{\Lambda d}^{}$, $a_{\Lambda\Lambda}^{}$, $g_1^{} (\Lambda_c)$, and 
$\Lambda_c$. 
The value of $\gamma_{\Lambda d}^{}$ can be fixed from the $\Lambda$ separation 
energy of the hypertriton and that of $a_{\Lambda\Lambda}^{}$ may be fixed from other 
experiments or possibly lattice QCD simulations.
However, there is no available experimental data to constrain the value of 
$g_1^{}(\Lambda_c)$.

When we do not introduce $g_1(\Lambda_c)$ in the theory, we obtain 
$a_1^{} \simeq - 25\sim -22$~fm with $\Lambda_c \simeq \Lambda_H$.
This may imply that the hypertriton-$\Lambda$ interaction is attractive 
but it is not strong enough to form a bound state.
Thus, if the ${}_{\Lambda\Lambda}^{\ \,4}$H bound state is formed, the main 
binding mechanism should stem from the mechanisms of high momentum region,
which is represented by the coupling $g_1^{} (\Lambda_c)$ in the present approach.
Therefore, to take into account this effect, we assume a formation of the 
${}_{\Lambda\Lambda}^{\ \,4}$H bound state and employ the results of 
the potential model calculations for the two-$\Lambda$ separation energy 
$B_{\Lambda\Lambda}$ for several values of $a_{\Lambda\Lambda}^{}$
to constrain the value of $g_1^{}(\Lambda_c)$. 
Using the fixed $g_1^{} (\Lambda_c)$ we then calculate $B_{\Lambda\Lambda}$
as a function of $a_{\Lambda\Lambda}^{}$. 
We also calculate the correlations between $B_{\Lambda\Lambda}$ and
$1/a_1^{}$, where $a_1^{}$ is the scattering length of the $S$-wave 
hypertriton-$\Lambda$ scattering for spin triplet channel.

As can be notably seen in the numerical results for the correlation between 
$B_{\Lambda\Lambda}$ and $a_{\Lambda\Lambda}^{}$ as given in 
Fig.~\ref{fig;BLLvsaLL}, when the cutoff is chosen to be the large scale of 
the theory, i.e., $\Lambda_c\simeq \Lambda_H$, $B_{\Lambda\Lambda}$ is 
insensitive to the value of $a_{\Lambda\Lambda}^{}$.
But, when $\Lambda_c$ is larger than $\Lambda_H$, say
$\Lambda_c \simeq 6 \Lambda_H$, $B_{\Lambda\Lambda}$ 
is sensitive to $a_{\Lambda\Lambda}^{}$, which gives results similar to 
the potential model predictions.
This would be a natural consequence because $a_{\Lambda\Lambda}^{-1}$
is a quantity of a large scale, $|a_{\Lambda\Lambda}^{-1}| \simeq
100 \sim 400$~MeV, compared to the typical scale of the system, 
$Q \sim \gamma_{\Lambda d}^{} \simeq 13.4$~MeV.
In addition, the dynamics that exhibits the sensitivity to the 
$\Lambda$-$\Lambda$ interaction above $\Lambda_c \simeq \Lambda_H$ 
is integrated out and its effect in high momentum is embedded in the 
contact interaction $g_1^{}(\Lambda_c)$.
Meanwhile, although the deuteron cluster theory with a large cutoff value
such as $\Lambda_c\simeq 6 \Lambda_H$ can reproduce the 
$a_{\Lambda\Lambda}^{}$-dependence of $B_{\Lambda\Lambda}$ similar to 
the potential model predictions, this would be inconsistent with the construction 
principles of EFT and it will miss the important dynamic mechanisms as discussed 
before. 
Therefore, the $a_{\Lambda\Lambda}^{}$-sensitivities in the physical 
observables for the ${}_{\Lambda\Lambda}^{\ \,4}$H hypernucleus, such as 
$B_{\Lambda\Lambda}$, inevitably depend on the scale of the theory. 
Investigating the $a_{\Lambda\Lambda}^{}$-sensitivity in more detail at another scale 
in the $_{\Lambda\Lambda}^{\ \,4}$H system requires to work with 
a non-cluster theory such as the pionless theory for four-body systems~\cite{PHM04}.

Experimentally, we still do not have enough information to judge whether the 
${}_{\Lambda\Lambda}^{\ \,4}$H system is bound or not.
This causes the difficulty for studying the energy levels of the 
${}_{\Lambda\Lambda}^{\ \,4}$H hypernucleus within EFT since the value of 
the contact interaction $g_1^{}(\Lambda_c)$ cannot be constrained by other 
information.
Therefore, it would be interesting to apply this approach to other 
double-$\Lambda$ hypernuclei, where some empirical data are available
such as the ${}_{\Lambda\Lambda}^{\ \ 6}$He system.
The ${}_{\Lambda\Lambda}^{\ \ 6}$He hypernucleus as a $\Lambda\Lambda \alpha$ 
three-body cluster system can be investigated in the scheme of EFT.
Because the binding energy, or equivalently
the two-$\Lambda$ separation energy, of $_{\Lambda\Lambda}^{\ \ 6}$He
is experimentally known, it can be used to determine the strength of the 
three-body contact interaction in the $\Lambda\Lambda\alpha$ system.
Moreover, because the $\alpha$ particle is more tightly bound than the deuteron,
the high momentum scale of the cluster theory becomes larger than $\Lambda_H$ 
of the present work.
Therefore, the study of $_{\Lambda\Lambda}^{\ \ 6}$He in Halo/Cluster EFT
can provide another tool to study $a_{\Lambda\Lambda}$ in the 
exotic systems and shed light on our understanding of strong interactions in the 
strangeness sector.
Work in this direction is under progress and will be reported elsewhere.

\acknowledgments

We are grateful to E.~Hiyama for valuable suggestions and discussions.
S.-I.A. thanks M.~Rho, Y.~Fujiwara, H.~Nemura, K.~Tanida, Y.-H. Song, and 
S.~W. Hong for fruitful discussions. 
We also acknowledge the warm hospitality of Asia Pacific Center for Theoretical 
Physics during the Topical Research Program and APCTP-WCU Joint Program.
The work of S.-I.A. was supported by the Basic Science Research Program through 
the National Research Foundation of Korea funded by the Ministry of Education 
(Grant No.~NRF-2012R1A1A2009430).
G.-S.Y. and Y.O. were supported by the National Research Foundation of Korea 
under Grant No.~NRF-2013R1A1A2A10007294.


\begin{thebibliography}{10}

\bibitem{RSY99}
\mbox{Th}. A.~Rijken, V.~G.~J. Stoks, and Y.~Yamamoto,
\newblock Phys. Rev. C \textbf{59}, 21 (1999).

\bibitem{PHM07}
H.~Polinder, J.~Haidenbauer, and U.-G. Mei{\ss}ner,
\newblock Phys. Lett. B \textbf{653}, 29 (2007).

\bibitem{FSN06}
Y.~Fujiwara, Y.~Suzuki, and C.~Nakamoto,
\newblock Prog. Part. Nucl. Phys. \textbf{58}, 439 (2007).

\bibitem{Jaffe77}
R.~L. Jaffe,
\newblock Phys. Rev. Lett. \textbf{38}, 195 (1977).

\bibitem{NPLQCD-10}
NPLQCD Collaboration, S.~R. Beane, E.~Chang, W.~Detmold, B.~Joo, H.~W. Lin,
  T.~C. Luu, K.~Orginos, A.~Parre{\~n}o, M.~J. Savage, A.~Torok, and
  A.~Walker-Loud,
\newblock Phys. Rev. Lett. \textbf{106}, 162001 (2011).

\bibitem{HALQCD-10}
HAL QCD Collaboration, T.~Inoue, N.~Ishii, S.~Aoki, T.~Doi, T.~Hatsuda,
  Y.~Ikeda, K.~Murano, H.~Nemura, and K.~Sasaki,
\newblock Phys. Rev. Lett. \textbf{106}, 162002 (2011).

\bibitem{DGPP63}
M.~Danysz \textit{et~al.\/},
\newblock Nucl. Phys. \textbf{49}, 121 (1963).

\bibitem{DDFM89}
R.~H. Dalitz, D.~H. Davis, P.~H. Fowler, A.~Montwill, J.~Pniewski, and J.~A.
  Zakrzewski,
\newblock Proc. Roy. Soc. Lond. A \textbf{426}, 1 (1989).

\bibitem{TAAA01}
H.~Takahashi \textit{et~al.\/},
\newblock Phys. Rev. Lett. \textbf{87}, 212502 (2001).

\bibitem{AAAB01}
J.~K. Ahn \textit{et~al.\/},
\newblock Phys. Rev. Lett. \textbf{87}, 132504 (2001).

\bibitem{FG02}
I.~N. Filikhin and A.~Gal,
\newblock Phys. Rev. Lett. \textbf{89}, 172502 (2002).

\bibitem{NAM02}
H.~Nemura, Y.~Akaishi, and K.~S. Myint,
\newblock Phys. Rev. C \textbf{67}, 051001(R) (2003).

\bibitem{Shoeb04}
M.~Shoeb,
\newblock Phys. Rev. C \textbf{69}, 054003 (2004).

\bibitem{NSAM04}
H.~Nemura, S.~Shinmura, Y.~Akaishi, and K.~S. Myint,
\newblock Phys. Rev. Lett. \textbf{94}, 202502 (2005).

\bibitem{SUB11}
B.~Sharma, Q.~N. Usmani, and A.~R. Bodmer,
\newblock Chinese Phys. Lett. \textbf{30}, 032101 (2013).

\bibitem{YAAF07}
KEK-PS E522 Collaboration, C.~J. Yoon \textit{et~al.\/},
\newblock Phys. Rev. C \textbf{75}, 022201(R) (2007).

\bibitem{GHH11}
A.~Gasparyan, J.~Haidenbauer, and C.~Hanhart,
\newblock Phys. Rev. C \textbf{85}, 015204 (2012).

\bibitem{STAR13}
STAR Collaboration, N.~Shah,
\newblock Nucl. Phys. \textbf{A904-905}, 443c (2013).

\bibitem{ExHIC-13}
ExHIC Collaboration, A.~Ohnishi \textit{et~al.\/},
\newblock Nucl. Phys. \textbf{A914}, 377 (2013).

\bibitem{RY06b}
\mbox{Th}. A.~Rijken and Y.~Yamamoto, nucl-th/0608074.

\bibitem{BV02c}
P.~F. Bedaque and U.~van Kolck,
\newblock Ann. Rev. Nucl. Part. Sci. \textbf{52}, 339 (2002).

\bibitem{BH04}
E.~Braaten and H.-W. Hammer,
\newblock Phys. Rep. \textbf{428}, 259 (2006).

\bibitem{JBKK73}
M.~Juric \textit{et~al.\/},
\newblock Nucl. Phys. \textbf{B52}, 1 (1973).

\bibitem{Griess05}
H.~W. Griesshammer,
\newblock Nucl. Phys. \textbf{A760}, 110 (2005).

\bibitem{KSW98a}
D.~B. Kaplan, M.~J. Savage, and M.~B. Wise,
\newblock Phys. Lett. B \textbf{424}, 390 (1998).

\bibitem{BHV99}
P.~F. Bedaque, H.~W. Hammer, and U.~van Kolck,
\newblock Nucl. Phys. \textbf{A676}, 357 (2000).

\bibitem{Wei79}
S.~Weinberg,
\newblock Physica \textbf{96A}, 327 (1979).

\bibitem{BKM95}
V.~Bernard, N.~Kaiser, and U.-G. Mei{\ss}ner,
\newblock Int. J. Mod. Phys. E \textbf{4}, 193 (1995).

\bibitem{BS00}
S.~R. Beane and M.~J. Savage,
\newblock Nucl. Phys. \textbf{A694}, 511 (2001).

\bibitem{AH04}
S.-I. Ando and C.~H. Hyun,
\newblock Phys. Rev. C \textbf{72}, 014008 (2005).

\bibitem{AH12}
S.-I. Ando and C.~H. Hyun,
\newblock Phys. Rev. C \textbf{86}, 024002 (2012).

\bibitem{Hammer01}
H.~W. Hammer,
\newblock Nucl. Phys.  \textbf{A705}, 173 (2002).

\bibitem{BV98a}
P.~F. Bedaque and U.~van Kolck,
\newblock Phys. Lett. B \textbf{428}, 221 (1998).

\bibitem{Ando13}
S.-I. Ando, arXiv:1302.7200.

\bibitem{PHM04}
L.~Platter, J.~Haidenbauer, and U.-G. Mei{\ss}ner,
\newblock Phys. Lett. B \textbf{607}, 254 (2005).

\end{thebibliography}
\end{document}